\documentclass[smallextended]{svjour3}
\usepackage{dcolumn}
\usepackage{bm}
\usepackage{times}
\usepackage{graphicx}

\begin{document}
\title{Current characteristics of mesoscopic rings in quantum Smoluchowski regime
\thanks{The work dedicated to Prof. Lutz Schimansky-Geier on the occasion of his 60th birthday. }}
\author{Sz. Rogozi\'{n}ski  \and L. Machura \and J. \L uczka \thanks{\email{Jerzy.Luczka@us.edu.pl}} }
\institute{Institute of Physics, University of Silesia, 40-007 Katowice, Poland }

\abstract{In normal mesoscopic metals of a ring topology persistent currents can be induced by 
threading the center of the ring with a magnetic flux. This phenomenon is an example of the famous 
Aharonov-Bohm effect. In the paper we study the current {\it vs} the external constant magnetic flux 
characteristics of the system driven by both the classical and the quantum thermal fluctuations. The 
problem is formulated in terms of Langevin equations in classical and quantum Smoluchowski regimes. 
We analyze the impact of the quantum thermal fluctuations on the current-flux characteristics. We 
demonstrate that the current response can be changed from paramagnetic to diamagnetic when the quantum 
nature of the thermal fluctuations increases. 
}
\maketitle

\section{Introduction}
Mesoscopic systems lay at the border between macroscopic and microscopic worlds.  Mesoscopic systems, 
consisting  of a large number of atoms,  are  too big to study their  properties by the quantum methods 
of individual atoms, and are  too small   to apply physical laws of the macro-world.   
For their description and modeling one should combine both methods  appreciating and recognizing  the 
role of quantum and classical  processes.  The appearance of a new length scale - the phase coherence 
length $l_c$ of electronic wave functions - introduces various regimes for   transport  phenomena 
influenced by quantum interference effects. 
The mesoscopic regime is characterized by small length scales and low temperatures. When the temperature
is lowered, the phase coherence length increases and the mesoscopic regime is extended to larger length scales. 
At sub-Kelvin temperatures, the  length scales are of the order of  micrometers. 
The most prominent mesoscopic effects  are:  the Aharonov-Bohm oscillations in the conductance of
mesoscopic structures, the quantum Hall effects, the universal conductance fluctuations and  
persistent currents in mesoscopic normal metal rings threaded by a magnetic flux. 
Persistent currents have been known to exist  in superconductors in which they are related to the existence 
of a state with zero resistance and the fact that a superconductor is a perfect diamagnet. 
The existence of persistent currents in normal (i.e. non-superconducting)  metals is, in fact, a manifestation
of the famous Aharonov-Bohm effect.  Persistent currents have been
predicted by Hund  in 1938 \cite{hund} and re-discovered later by others \cite{but}.  
Inspired by these findings, the first experiment was performed in the early 1990s by measuring the 
magnetization of an array of about ten million not connected micron-sized copper rings  \cite{buh}.
Other experiments also supported existence of persistent currents \cite{chandra}. 
We should also  recall recent definitive measurements of persistent currents in nanoscale gold and aluminium rings
\cite{harris,bluhm}.  The team \cite{harris}  has developed a new technique for detecting persistent currents that 
allows  to measure the persistent current  over a wide range of temperatures, ring sizes, and magnetic 
fields.  They have  used nanoscale cantilevers, an entirely novel approach to indirectly measure the current through 
changes in the magnetic force it produces as it flows through the ring.
The second team \cite{bluhm} has studied thirty  three individual rings, in which they have  employed a scanning 
technique (a SQUID microscope).  
The rings are very small, each only between one and two micrometers  in diameter and 140 nanometers  thick. They 
are made of high-purity gold. Each was scanned individually, unlike past experiments on persistent currents 
conducted by other groups. In total they were scanned approximately 10 million times.
Both works mark the first time that the theory has been {\it experimentally proven} to a high degree.

In our previous papers \cite{rogo} we have proposed a two-fluid model for the dynamics of the magnetic flux that 
passes through a mesoscopic ring. It is described in terms of an ordinary differential equation with an additional 
random force. It is analogous to the well known model of a capacitively and a resistively shunted Josephson 
junction \cite{barone}. The classical part consists of 'normal' electrons carrying dissipative current. The 
quantum part is formed by those electrons which maintain their phase coherence around the circumference of the ring 
(it is a counterpart of the Cooper pairs of the electrons in the superconducting systems). The effective dynamics 
is than determined by a classical Langevin equation \cite{lutz} with a Johnson noise describing classical thermal 
equilibrium fluctuations. For low temperatures, quantum nature of thermal fluctuations should be taken into account. 
To this aim we apply the approach based on the so called quantum Smoluchowski equation as introduced in 
Ref. \cite{ankerhold1} and in other versions in the following Refs. \cite{luczka,ankerhold2,coffey}. 
 
The paper is organized as follows. First, in the Sec. 2, we present a model of capacitively and 
resistively shunted Josephson junction in order to demonstrate readers the analogy between both models. 
Next, in the Sec. 3, we briefly present our model for the flux dynamics in the normal metal rings. 
In the Sec. 4, we define the quantum Smoluchowski regime following by the Sec. 5, presenting the 
dimensionless variables and parameters. In the Sec. 6, we study the current characteristics in the 
stationary states for both classical and quantum Smoluchowski domain. 
We end this work with the summary and conclusions.  

\section{Superconducting ring}
For clarity of modeling the current characteristics in non-superconducting rings, we present the  
well-known approach to describe a quasi-classical regime of superconducting rings. To this aim, 
let us consider a superconducting loop (ring, cylinder, torus)  interrupted by a Josephson junction.  
This element is a basic unit of various SQUID devices. The phase difference 
$\psi$ of the Cooper pair wave function across the junction is related to the magnetic flux  
$\phi$ threading the ring via the relation \cite{barone}
\begin{equation}\label{relat}
\psi =2\pi (n-\phi/\phi_0), 
\end{equation}
where $2\pi n$  is the phase change per cycle around the ring and $\phi_0=h/2e$ is the flux quantum. 
When the external magnetic field is applied, the  total flux is 
\begin{equation}\label{L}
\phi =\phi_e + L I,  
\end{equation}
where $\phi_e$ is the flux generated  by an applied external magnetic  field, $L$ is the self-inductance of the ring and  
$I$ is the total  current flowing in the ring.  We model the Josephson element  in terms of the resistively and capacitively shunted junction for which the current  consists of  three components \cite{kos}, namely, 
\begin{equation}\label{IJ}
I = I_C + I_R  + I_J = \frac{ \phi - \phi_e}{L},   
\end{equation}
where $I_C$  is a displacement current accompanied with the junction capacitance $C$,  $I_R$ is a normal  (Ohmic) current characterized by the normal state resistance $R$ and $I_J$ is the Josephson supercurrent.  In the right-hand side, the relation (\ref{L}) has beed used.

Combining Eqs. (\ref{relat})-(\ref{IJ}) with 
 the  second Josephson  relation $d\psi/dt=2eU/\hbar$, where $U$ is the voltage drop across the junction, 
we get the Langevin-type equation in the  form \cite{barone}
\begin{equation}\label{JJ}
C\frac{d^2 \phi}{dt^2}+\frac{1}{R}\frac{d \phi}{dt} + I_0 \sin \phi =- \frac{ \phi - \phi_e}{L} 
+\sqrt{\frac{2k_BT}{R}}\Gamma(t). 
\end{equation}
%
This  equation  has a  mechanical interpretation: it can describe  the  'position' $\phi(t)$  of 
the  Brownian particle  moving in the washboard potential 
\begin{equation}\label{C}
W(\phi)= \frac{(\phi -\phi_e)^2}{2L} - I_0 \cos \phi. 
\end{equation}
The first term originates from the external bias and the self--inductive interaction of the magnetic flux whereas the second term is the  supercurrent modified by the quantum flux.   
The ubiquitous thermal equilibrium noise $\Gamma(t)$ 
consists of  Johnson noise associated with the resistance $R$. 
 The parameter $k_B$
denotes the Boltzmann constant and $T$ is  temperature of the
system. The Johnson noise  is  modeled by
$\delta$-correlated Gaussian white noise of zero mean,  $\langle\Gamma(t)\rangle = 0$,   and
unit intensity, i.e.,  $\langle\Gamma(t) \Gamma(u)\rangle = \delta(t-u)$.

\section{Normal metal ring}  

Now, let us consider a non-superconducting ring. When the external magnetic field is applied, the actual flux is given by 
\begin{equation}\label{LL1}
\phi =\phi_e + L I,  
\end{equation}
where $\phi_e$ is the flux generated  by an applied external magnetic  field, $L$ is the self-inductance of the ring  and  
$I$ is the total  current flowing in the ring.  At zero temperature,
the ring displays 
persistent and non-dissipative  currents $I_{P}$ run by  phase-coherent
electrons.  It is analogue of the Josephson supercurrent $I_J$. 
At  non-zero temperature,  a part of electrons
becomes 'normal' (non-coherent) and the amplitude of the
persistent current decreases. Moreover, 
 resistance of the ring and thermal fluctuations  should be taken into account.   Therefore for temperatures  $T>0$,  the total current consists of three parts, namely,
\begin{equation} \label{I}
I=I_C + I_R + I_{P} =  \frac{ \phi - \phi_e}{L}. 
\end{equation}
 What we need is  the expression  for the  persistent 
current $I_{P}$. It is s function of the magnetic flux $\phi$  and depends on the parity of
the number of coherent electrons.  Let $p$ denotes the
probability of an even number of coherent electrons and  $1-p$ is the  probability of an odd number of coherent electrons. Then the  persistent current  can be expressed  in the form  \cite{cheng}
\begin{eqnarray}
I_{P}=I_{P}(\phi)=p\,I_{E}(\phi)+(1-p)\,I_{O}(\phi),
\end{eqnarray}
where
\begin{eqnarray} \label{persi}
I_{E}(\phi)=I_{O}(\phi +\phi_0/2) 
=I_0\sum_{n=1}^\infty A_n(T/T^*) \cos(nk_F l) \sin ( 2n\pi \phi /\phi _0), 
\end{eqnarray}
where   $I_0$ is the maximal current at zero temperature.
 The temperature dependent amplitudes are determined by the relation 
\cite{cheng}
\begin{eqnarray}
A_n(T/T^*)= \frac{4T}{\pi T^*}\frac{\exp(-nT/T^*)}{1-\exp(-2nT/T^*)}
,
\end{eqnarray}
 where the  characteristic temperature $T^*$ is  proportional to the
energy gap $\Delta _F$ at the Fermi surface,  $k_F$ is the 
Fermi momentum and $l$ is the circumference of the ring. 
If the number $N$ of electrons is fixed then $k_F=\pi N/l$ and the persistent current takes the form 
\begin{eqnarray} \label{per2}
I_{P}(\phi) =I_0\sum_{n=1}^\infty A_n(T/T^*)\sin ( 2n\pi \phi /\phi _0) [p + (-1)^n (1-p)].
\end{eqnarray}
As a result, from Eq. (\ref{I})  we obtain the  equation of motion in the form \cite{rogo}
\begin{eqnarray}\label{PC}
C\frac{d^2\phi}{dt^2} + \frac{1}{R}\frac{d\phi}{dt} = 
-\frac{1}{L}(\phi-\phi_{e}) + I_{P}(\phi)
+ \sqrt{\frac{2 k_BT}{R}}\;\Gamma (t) \nonumber\\ 
 = - \frac{dV(\phi)}{d\phi} + \sqrt{\frac{2 k_BT}{R}}\;\Gamma (t),
\end{eqnarray}
where the  "potential" $V(\phi)$ reads 
\begin{eqnarray}\label{W}
V(\phi)&=&\frac{1}{2L} \left(\phi - \phi_e\right)^2 
+ \phi_0 I_0\sum_{n=1}^{\infty}
\frac{A_n(T/T^*)}{2n\pi} \cos\left(2n\pi \frac{\phi}{\phi_0}\right) \;[p+(-1)^n (1-p)]. 
\end{eqnarray}
In the above Langevin equation,  $C$  and $L$ are, respectively, the capacitance  and inductance of the ring.  
It was shown  \cite{kopietz}, that  the  energy associated with long-wavelength 
and low-energy charge fluctuations is determined by classical charging 
energies and therefore the ring  behaves as it were a classical capacitor. 
 The flux dependence of these energies yields the contribution 
to the persistent current  \cite{capac1}. 
The capacitance  becomes essential if the ring accommodates  a stationary impurity or a quantum dot \cite{dot}. 
 Moreover, in the mesoscopic domain the standard description
of a capacitor in terms of the geometric capacitance  
(that relates the charge  on the plate to the voltage 
across the capacitor) gives way to a more complex notion of capacitance  which depends on the properties of  conductors  \cite{butikerC}.   

Note that Eqs. (\ref{JJ}) and (\ref{PC}) have a similar structure.  The  difference is not only in the form
of the potential but also in  temperature  dependence of the potential in the case the normal metal rings.


\section{Quantum Smoluchowski regime}

In both models (for superconducting and non-superconducting rings), thermal equilibrium fluctuations are modeled  as classical fluctuations of zero correlation time. When temperature is lowered, quantum nature of  fluctuations starts to  play a role,   fluctuations become correlated  and leading quantum corrections should be taken into account.  It is not a simple task and a general method how to incorporate quantum corrections in a  case described by Eq. (\ref{PC}) is not known.  
However, in the quantum Smoluchowki regimes  \cite{ankerhold1}, where the charging effects  (related to the capacitance $C$) can be neglected,  the system can be described by the "overdamped" Langevin equation -  the so-named quantum Smoluchowski equation 
\cite{ankerhold1,luczka}.  For a Brownian particle it corresponds to neglecting inertial effects related to the mass of a particle.  The quantum Smoluchowski equation  has the same structure as a classical Smoluchowski equation, in which the  diffusion coefficient $D_0=k_BT/R$ is  modified due to quantum effects like tunnelling, quantum reflections and purely quantum fluctuations.   In terms of  the Langevin equation (\ref{PC}), it assumes the form
\begin{equation}\label{QOV}
\frac{1}{R}\frac{d\phi}{dt} = - \frac{dV(\phi)}{d\phi} +
\sqrt{2D_{ \Lambda}(\phi)}\;\Gamma (t). 
\end{equation}
This equation has to be interpreted  in the Ito sense \cite{gard}.  
The  modified diffusion coefficient $D_{ \Lambda}(\phi)$ takes the form \cite{luczka} 
\begin{eqnarray}
D_{ \Lambda}(\phi)=\frac{D_0}{1-\Lambda V''(\phi)/k_BT}, 
\end{eqnarray}
where the prime denotes differentiation with respect to the argument of the function.  
The quantum correction is  characterized by  the parameter 
$\Lambda$. It measures a deviation of the quantal flux fluctuations from 
its classical counterpart, namely, 
\begin{equation}\label{Lam1}
\Lambda=\langle \phi^2\rangle_{Q} -\langle \phi^2\rangle_{C},
\end{equation}
where $\langle \cdots \rangle$ denotes equilibrium average, the subscripts 
$Q$ and $C$ refer to quantal and classical cases, respectively. 
Let us determine the range of applicability of the quantum Smoluchowski regime. The classical Smoluchowski limit corresponds to the case when charging effects can be neglected. Formally,  
one can put $C=0$  in the inertial term of Eq. (\ref{PC}), which is 
related to the strong damping limit of the Brownian particle. In the
 case studied here it means that \cite{ankerhold1}
\begin{equation}\label{ineq1}
\omega_0 CR \ll 1, 
\end{equation}
where the  frequency $\omega_0$ is a typical frequency of the bare system 
and its inverse corresponds to a characteristic time of the system.  In such a case,    Eq. (\ref{Lam1}) takes the form \cite{rogo}
\begin{equation}\label{Lam3}
\Lambda = \frac{ \hbar R}{\pi}\left[ \gamma + 
\Psi\left(1+\frac{\hbar}{2\pi CR k_BT}\right)  \right],
\end{equation}
where   the psi function $\Psi(z)$ is the logarithmic derivative of the 
Gamma function and $\gamma \simeq 0.5772$ is the Euler constant. 

The separation of time scales, on which the flux relaxes and the conjugate observable (a charge) is already equilibrated, requires the second  condition, namely,  
\begin{equation}\label{ineq2}
\omega_0 CR \ll k_BT/\hbar \omega_0.
\end{equation}
 In the deep quantum regime, i.e. when 
\begin{equation}\label{ineq3}
 k_BT \ll \frac{\hbar}{2\pi CR},
\end{equation}
 the correction  parameter (\ref{Lam3}) simplifies to the form  
\begin{equation}\label{ln}
\Lambda=\frac{\hbar R}{\pi}\left[\gamma+\ln\left(\frac{\hbar}{2\pi CR k_BT}\right)\right]. 
\end{equation}
In order to identify precisely the  quantum Smoluchowski regime, we have to determine a typical frequency $\omega_0$ or the corresponding characteristic time $\tau_0 \propto 1/\omega_0$. There are many characteristic 
times in the system, which can be explicitly extracted from the 
evolution equation (\ref{PC}), e.g.  $CR$, $\hbar/ k_BT$,  $\phi_0/(RI_0)$.  The characteristic time 
$\tau_0 = L/R$ is the inductive  time of the ring and  for a typical
mesoscopic ring, $L/R$  is in the picosecond range. 
Therefore, in the quantum Smoluchowski regime, 
all the above inequalities 
(\ref{ineq1}), (\ref{ineq2}) and (\ref{ineq3}) should be fulfilled 
for $\omega_0 =2\pi/\tau_0$. 
Because the diffusion coefficient cannot be negative,    the parameter $\Lambda$ should be chosen  small enough to satisfy 
the condition $D_{\Lambda}(\phi)\ge 0$ for all values of $\phi$.  

The "overdamped" Langevin equation (\ref{QOV}) describes a classical 
Markov stochastic process and its probability density $P(\phi, t)$ obeys the Fokker-Planck equation 
\cite{gard}, namely, 
\begin{equation}\label{FP}
\frac{1}{R} \, \frac{\partial}{\partial t} P(\phi, t) = \frac{\partial}{\partial \phi} \left[
\frac{dV(\phi)}{d\phi} P(\phi, t) \right] + \frac{\partial^2}{\partial \phi^2} 
 \left[ D_{\Lambda}(\phi) P(\phi, t) \right] . 
\end{equation}
We wish to analyze an  averaged  stationary current  $\langle I \rangle$ flowing in the ring which can be obtained from Eq. (\ref{I}): 
\begin{equation} \label{Ia}
\langle I \rangle =  \frac{1}{L} \left[ \langle \phi \rangle - \phi_e \right], 
\end{equation}
where the averaged  stationary magnetic flux $ \langle \phi \rangle$ is calculated from the equation 
\begin{eqnarray} 
\label{phia}
 \langle \phi \rangle = \int_{-\infty}^{\infty} \phi \; P(\phi) d \phi,  \quad 
 P(\phi) = \lim_{t \to \infty} P(\phi, t), 
\end{eqnarray}
where $P(\phi)$ is a stationary probability density, 
\begin{eqnarray}\label{Ps}
P(\phi)= C_0  D^{-1}_{\Lambda}(\phi) 
\exp\left[- U(\phi)\right], 
\end{eqnarray}
where $C_0$ is the normalization constant and the generalized thermodynamic potential  $U(\phi)$ reads
\begin{eqnarray}\label{thermo}
U(\phi) =\int \frac{dV(\phi)}{d\phi} {D^{-1}_{\Lambda}(\phi)}\;d\phi.
\end{eqnarray}
Because the potential $V(\phi)$ depends on the external flux $\phi_e$,  the  averaged stationary  
current   (\ref{Ia}) is a non-linear function  of $\phi_e$.  Eqs. (\ref{Ia}) - (\ref{thermo}) 
form a closed set of equations from which the current characteristics 
$\langle I \rangle = f(\phi_e)$  as a certain function $f$ of the external magnetic  flux  $\phi_e$ can be obtained.  It is an analogous of the current-voltage characteristics  for electrical circuits.


\section{Dimensionless variables and parameters}
To analyze the current-flux characteristics in the  stationary  state,  
we first  introduce dimensionless variables and parameters.  
The rescaled flux $x=\phi/\phi_0$.  
Then Eq. (\ref{Ia}) can be rewritten in the dimensionless form  
\begin{equation}\label{LL}
i = \langle x \rangle -x_e, \quad i= \langle I \rangle L/\phi_0, \quad x_e=\phi_e/\phi_0, 
\end{equation}
where $i, \langle x \rangle$ and $x_e$ are  dimensionless averaged current, averaged flux and external flux, respectively.  

The stationary probability density $p(x)$  takes the
\begin{eqnarray}\label{ps}
p(x)= N_0  D^{-1}(x) \exp\left[-\Psi_{\lambda}(x)\right], 
\end{eqnarray}
where $N_0$ is the normalization constant and 
the generalized thermodynamic potential  reads
\begin{equation} 
\label{Psi}
\Psi_{\lambda}(x) =\int\frac{dV(x)}{dx} \,D^{-1}_{\lambda}(x) \,dx.
\end{equation}
The rescaled potential reads 
\begin{eqnarray} \label{V(x)}
V(x)=\frac{1}{2}(x-x_e)^2 + B(x),
\end{eqnarray}
where 
\begin{eqnarray}
  \label{B}
 B(x)= \alpha \sum_{n=1}^{\infty}\frac{A_n(T_0)}{2n\pi} \cos(2n\pi x) [p + (-1)^n (1-p)]
\end{eqnarray}
with the dimensionless temperature $T_0=T/T^*$ and $\alpha = LI_0/\phi_0$.  
The rescaled modified diffusion function $D_{\lambda}(x)$  assumes the form 
\begin{eqnarray}
\label{D(x)}
D_{\lambda}(x)=\frac{\beta^{-1}} {1-\lambda\beta V''(x)}   
\end{eqnarray}
with  $\beta^{-1}= k_BT/2E_m = k_0 T_0$,   the 
elementary magnetic flux energy $E_m=\phi_0^2/2L$ and 
$k_0= k_BT^*/2E_m$ is the ratio of two characteristic energies.  
The dimensionless quantum correction parameter 
\begin{equation}\label{lam}
\lambda =  \lambda_0 \left[ \gamma + 
\Psi\left(1+   \frac{\epsilon}{T_0}\right)  \right],
\quad
\lambda_0=\frac{ \hbar R}{\pi\phi_0^2}, 
\quad
\epsilon = \frac{\hbar/2\pi CR}{k_BT^*}.
\end{equation}
 Remember that the Smoluchowski regime corresponds to the strong coupling limit.  For classical systems, i.e. when the quantum correction  parameter $\lambda =0$,  the stationary state  is  a Gibbs state,  i.e.  
$p(x) \propto \exp[-\beta V(x)]$.   For quantum systems,   due to the $x$-dependence of the modified diffusion coefficient 
$D_{\lambda}(x)$,  the stationary state (\ref{ps})   is not  a Gibbs state. 
However,  it  is a thermal equilibrium state. 

\begin{figure}[htbp]
  \begin{center}
    \includegraphics[width=0.49\textwidth]{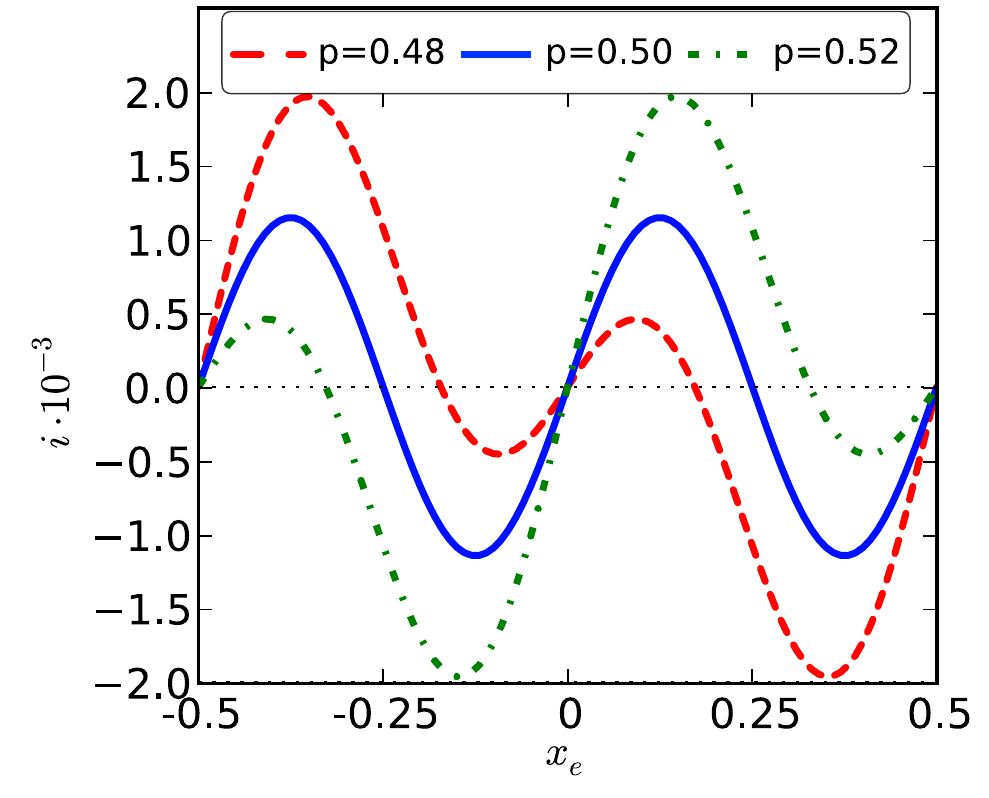}
  \end{center}
  \caption{(color online) 
    The dimensionless current $i$  {\it vs} the external magnetic flux $x_e$ for three values of the 
    probability $p$ of an even number of coherent electrons in the ring. 
    Please note that one can obtain the persistent  current for slightly asymmetric case $p=0.48$ just by shifting
    the presented characteristic for $p=0.52$ by $ x_e = 0.5 $.
    It is the case of "classical" thermal fluctuations, i.e. when the  quantum correction parameter $\lambda_0 =0$.
    Other dimensionless parameters are: $\alpha = 0.1$, $T_0 = 0.5$, $k_0 = 0.08$.  
  }
  \label{fig1}
\end{figure}

\section{Current -- flux characteristics in stationary states}\label{current-flux}

A persistent current is a periodic function of the magnetic flux with a period given by a single-electron  flux unit  $\phi_0=h/e$.  To what extent the current  is highly sensitive to a variety of subtle effects such as an electron--electron interaction, defects, disorder, coupling to an environment and other degrees of freedom, it is still a topic of controversy and persistent discussion. 
Fortunately,  novel techniques developed recently such as the  microtorsional magnetometer \cite{harris} 
and  scanning SQUID \cite{bluhm} allow to measure the persistent current in metal rings over a wide range of magnetic fields, temperatures and ring sizes.  Like many mesoscopic effects, the persistent current in real systems depends on the particular realization of disorder and thus varies between nominally identical rings, cf. the term $\cos(nk_Fl)$ in  Eq. (\ref{persi}) which in practice is random.  
In Fig. 1 we depict   the  well-known dependence of current  upon  the external magnetic flux $x_e$ for three selected  values of the probability $p$ of an even number of coherent electrons in the ring.  The choice of  values of $p$ is arbitrary but shapes of the current are similar to those observed in experiments. 

We now focus on the influence of quantum thermal fluctuations on persistent currents. 
The deviation from the "classicality"  is measured by the  dimensionless parameter  $\lambda$ which depends on $\lambda_0$ (the material constant) and temperature, 
see Eq. (\ref{lam}).  It is instructive to compare basic quantities characterizing  the system. 
In Fig. 2 we show the  generalized thermodynamic potential $\Psi_{\lambda}(x)$, the modified diffusion 
function $D_{\lambda}(x)$ and the stationary probability density $p(x)$ for two values of the quantum 
correction strength $ \lambda_0 $.  Three panels (a), (b) and (c) are presented for the case $x_e =0$ (the vanishing  
external flux). The case  $\lambda_0 =0$ corresponds  to classical thermal fluctuations and 
$\Psi_{0}(x)$ is a bare potential $V(x)/k_0 T_0$.  We  note  that the  generalized thermodynamic 
potential $\Psi_{\lambda}(x)$ for various $\lambda$  changes  only slightly. On the contrary, the  
state-dependent diffusion function  is a periodic function of the magnetic flux  and possess maxima 
and minima. It is a radical difference to the classical case $\lambda =0$ for which $D_0(x)=D_0=k_0T_0$ 
is  a constant function (thin solid blue line  and thin dashed red line in panel (c)).  The maxima of $D_{\lambda}(x)$ can be interpreted as 
a higher effective local temperature.  They are located  at $x_e=1/4 \; \mbox{mod}(1/2)$. 
The impact of quantum corrections on the stationary probability density $p(x)$  seems to be rather
insignificant. One can observe  a small deformation  around the peak of the density: for lower temperature  and non-zero  $\lambda_0$  
 the peak becomes slightly higher and narrower and the tails do not diverge in the quantum
case as fast as in the classical one.  

\begin{figure}[htbp]
  \begin{center}
    \includegraphics[width=0.99\textwidth]{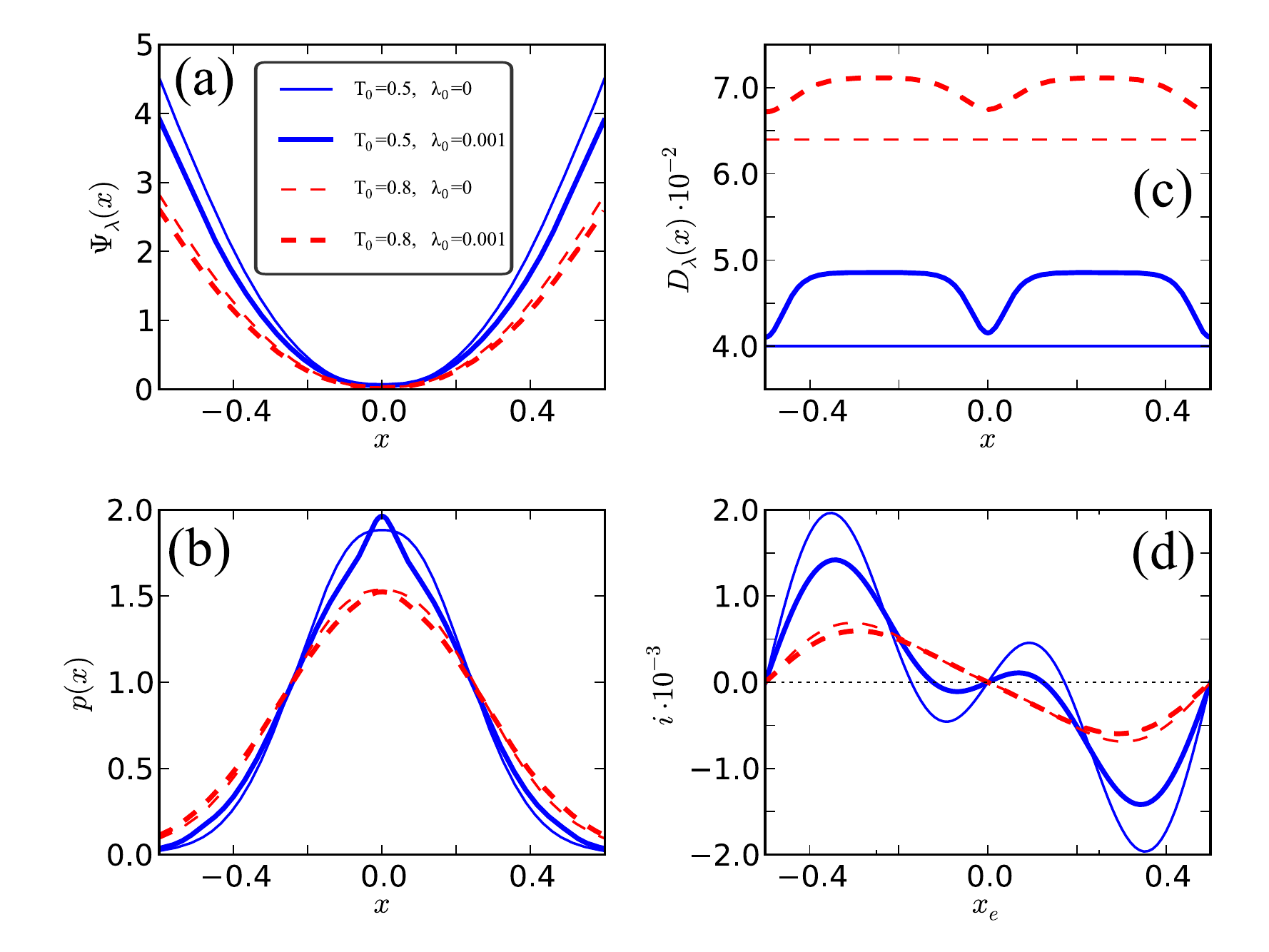}
  \end{center}
  \caption{(color online) 
  Four characteristics of the normal metallic ring in the absence (panels a -- c) and presence (panel d) of
  the external flux are demonstrated for two values of the dimensionless temperature: $ T_0 = 0.5, 0.8 $ and for two different values of the quantum 
  fluctuations strength:  $ \lambda_0 = 0, 0.001 $. The key for reading this plot is as the following: 
  blue (solid) lines denote curves corresponding to the lower temperature $T_0 = 0.5 $ and the red (dashed) lines correspond to the higher temperature $ T_0 = 0.8 $; 
  the thin lines correspond to $ \lambda_0 = 0 $ (classical thermal fluctuations) and the thick lines correspond  to   $ \lambda_0 = 0.001 $ (quantum thermal fluctuations).   
  In panel (a) we present the thermodynamic potential defined in (\ref{Psi}). In the classical 
  case ($ \lambda_0 = 0 $) it reduces to $ V(x) / k_0 T_0 $. One can notice only small deviations from the classical case when 
  the rescaled parameter $ \lambda_0 $ is increased to $ 0.001 $. On the panel (b) we illustrate  the 
  corresponding stationary probability density function $ p(x) $. The bell -- shaped curve in the 
  classical limit is slightly deformed for non-zero $ \lambda_0 $. The peak tends to be narrower and reach 
  higher values and the tails decay slower as we increase $ \lambda_0 $. This effect is stimulated collectively 
  by the  thermodynamic potential and the effective diffusion presented on panels (a) and (c), respectively.
  Please note that for the classical case,  the flux dependence of  $ D_\lambda(x) $ disappears and constantly 
  equals $ k_0 T_0 = 0.04 $  for $ T_0 = 0.5 $ and $ 0.064 $ for  $ 0.8 $.   
  The most significant influence of the temperature is found in panel (d), where the current--flux characteristics are  displayed. For small external load,
  around flux $ x_e = 0 $, the system responses in a completely different way for  two selected 
  temperatures. For $ T_0 = 0.5 $ in both classical and quantum cases the persistent current is paramagnetic. 
  If we, however, increase the temperature to $ T_0 = 0.8 $, the situation changes drastically and the   susceptibility for this higher temperature becomes diamagnetic. 
  Other rescaled parameters are set as the following 
  $\alpha = 0.1$, $k_0 = 0.08$, $p = 0.48$, $\epsilon = 100$.
  }
  \label{fig2}
\end{figure}
Finally, we  analyze the influence of quantum thermal fluctuations on the current-flux characteristics $i=i(x_e)$. 
It is illustrated in panel (d) of Fig. 2 and in Fig. 3.   
Two solid blue lines  in panel (d)  of Fig. 2 are qualitatively  similar to the experimental curve 
shown in figure S6(A) in  the Supporting Online Material \cite{supp} of the paper \cite{harris}. 
We observe that in all cases of  quantum thermal fluctuations the amplitude of persistent currents is reduced in comparison to classical thermal fluctuations  case ($\lambda_0=0$) both in the paramagnetic regime ($T_0=0.5$) and diamagnetic regime ($T_0=0.8$). 
The parameter regime depicted in Fig. 3 is much more interesting.  For $T_0=0.6$, 
in the "classical"  case, the persistent current is {\it paramagnetic}, i.e. $i= \eta x_e$  with the 
positive slope  $\eta > 0$ in the vicinity of $x_e =0$. It is a linear response regime where the transport coefficient (susceptibility) 
$\eta = \lim_{x_e \to 0} [i(x_e)/x_e]$.  
%
\begin{figure}[htbp]
  \begin{center}
    \includegraphics[width=0.49\textwidth]{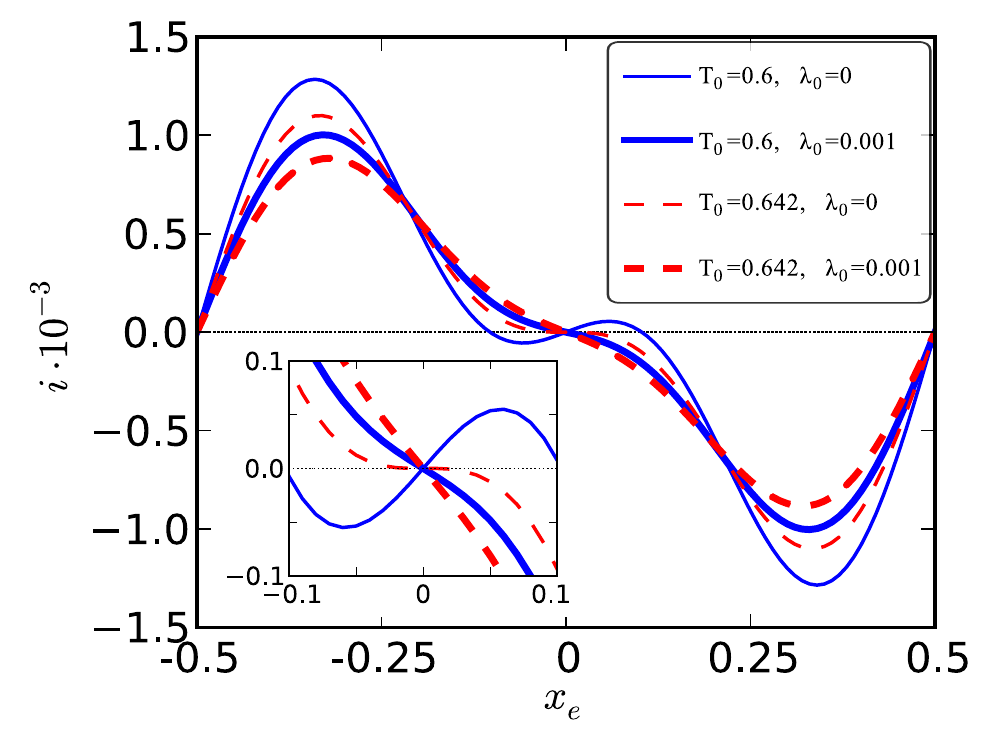}
  \end{center}
  \caption{(color online) 
  The dimensionless current $i$ {\it vs} the external magnetic flux $x_e$ for two values of the dimensionless temperature 
  $ T_0 = 0.6, 0.642 $ and two values of the parameter $ \lambda_0 = 0$ (classical thermal fluctuations) and 
  $ 0.001 $ (quantum thermal fluctuations). Again, the key for reading this plot is as for the previous figure: 
  blue (solid) lines denote curves corresponding to the lower temperature $ T_0 = 0.6 $ and the red (dashed) lines correspond to the higher temperature $ T_0 = 0.642 $; 
 the thin lines correspond to $ \lambda_0 = 0 $ and thick lines corresponds  to 
  $ \lambda_0 = 0.001 $.  The most significant influence of the 'quantum parameter' $ \lambda_0$ 
  is found for the rescaled temperature $ T_0 = 0.6 $, where the current $ i $ changes its behavior
  from paramagnetic to diamagnetic one  just by adjusting $ \lambda_0 $ from $ 0 $ to $ 0.001 $. Moreover,
  for the presented set of the system parameters we can find characteristic cross-temperature at 
  $ T_0 = 0.642 $ where the magnetic susceptibility is zero in the classical case and diamagnetic in 
  quantum,  see red (dashed) line for details. 
  Rescaled parameters are set as the following $\alpha = 0.1$, $k_0 = 0.08$, $p = 0.48$, $\epsilon = 100$.
  }
  \label{fig3}
\end{figure}
 If temperature is a litlle bit higher ($T_0=0.642$),  the susceptibility $\eta =0$  zero in the classical case. 
If  quantum corrections are taken into account,    the susceptibility $\eta < 0$ ,  the  slope of the $i-x_e$ curve in negative and the current becomes diamagnetic. 
 The most interesting observation is  that the persistent current can change its character from the paramagnetic to diamagnetic phase and the sign of the low-field magnetic response  depends on the level of the quantum corrections.  Our detailed numerical analysis shows 
that the  sign of  magnetic susceptibility  can easily be affected by system parameters and therefore is not robust against small perturbations.  This is  what has been observed in many experiments regarding the 
paramagnetic or/and diamagnetic persistent currents. The best illustration of what we state here is the  response of 15 nominally identical ring presented in Fig.  2 in Ref. \cite{bluhm}:  e.g. for the ring 1 the current is paramagnetic while for the ring 2  it is diamagnetic. 

\section{Conclusions}

In many cases and for various systems at the "intermediate" temperatures, the semi-classical theory is insufficient and 
the quantum corrections should be involved. It has been shown in the literature that in the strong friction 
limit, the quantum effects are restricted not only to low temperatures and therefore they should be incorporated 
for the higher temperatures as well. This is so because the quantum fluctuations, even if reduced for one variable, 
are enlarged for the conjugate variable. 
The dynamics as well as the stationary states in this regime can be modeled by the quantum
Smoluchowski equation. In other words, the quantum non-Markovian stochastic process is  
approximated by the classical Markovian process with the modified, state-dependent diffusion 
function. 

The role of the quantum corrections on the current-flux characteristics is addressed in this work.  
A general conclusion is that the quantum thermal fluctuations reduce the amplitude of the persistent currents: 
the current amplitude is always smaller than the corresponding "classical" one. In the quantum case, the diffusion 
constant becomes a periodic function of the magnetic flux. Maxima and minima of the diffusion function can be 
interpreted in terms of the higher and lower local temperature. There are parameters regimes where the system 
response changes the character from the paramagnetic to diamagnetic, when the quantum effects of thermal 
fluctuations increases. It would be interesting to extend the current study by including the time-dependent 
drivings modeled by the time-periodic magnetic fields. One could expect novel transport phenomena like a negative 
susceptibility which for Brownian particles corresponds to negative mobility \cite{mach} or negative 
conductances \cite{kos}. 

\section*{Acknowledgment}
The work supported by the ESF Program 
{\it Exploring the Physics of Small Devices}.  J. {\L}. wishes to thank Lutz Schimansky-Geier for 
long-term  friendship, hospitality and collaboration. Sto lat, Lutz!

\end{document}